\documentclass[conference]{IEEEtran}
\IEEEoverridecommandlockouts
\usepackage{amsmath,amssymb,amsfonts}
\usepackage{algorithmic}
\usepackage{textcomp}
\usepackage{xcolor}
\usepackage{flushend}
\usepackage{float}
\usepackage{comment}
\usepackage{dirtytalk}
\usepackage{algorithm} 
\usepackage{algorithmic}  
\usepackage[algo2e]{algorithm2e}

\def\BibTeX{{\rm B\kern-.05em{\sc i\kern-.025em b}\kern-.08em
    T\kern-.1667em\lower.7ex\hbox{E}\kern-.125emX}}
\PassOptionsToPackage{bookmarks={false}}{hyperref}      %to disable bookmarks

\usepackage{color,soul}         %HIGHLIGHT
\usepackage{cite}                                    %Multiple citations
\usepackage[pdftex]{graphicx,hyperref,color}      %Multiple citations
\begin{document}
\title{Automatic and Flexible Transmission of Semantic Map Images using Polar Codes for End-to-End Semantic-based Communication Systems}
\author{ \IEEEauthorblockN{{Hossein~Rezaei, Thushan~Sivalingam, Nandana~Rajatheva}}
\IEEEauthorblockA{Centre for Wireless Communications,~University of Oulu, Finland\\
E-mail: \{hossein.rezaei, thushan.sivalingam, nandana.rajatheva\}@oulu.fi}}
\maketitle
%\vspace*{-0.3pt}
\begin{abstract}
Semantic communication represents a promising roadmap toward achieving end-to-end communication with reduced communication overhead and an enhanced user experience. The integration of semantic concepts with wireless communications presents novel challenges. This paper proposes a flexible simulation software that automatically transmits semantic segmentation map images over a communication channel. An additive white Gaussian noise (AWGN) channel using binary phase-shift keying (BPSK) modulation is considered as the channel setup. The well-known polar codes are chosen as the channel coding scheme. The popular COCO-Stuff dataset is used as an example to generate semantic map images corresponding to different signal-to-noise ratios (SNRs). To evaluate the proposed software, we have generated four small datasets, each containing a thousand semantic map samples, accompanied by comprehensive information corresponding to each image, including the polar code specifications, detailed image attributes, bit error rate (BER), and frame error rate (FER). The capacity to generate an unlimited number of semantic maps utilizing desired channel coding parameters and preferred SNR, in conjunction with the flexibility of using alternative datasets, renders our simulation software highly adaptable and transferable to a broad range of use cases.
\end{abstract}
\begin{IEEEkeywords}
end-to-end communication, error-correcting codes, polar code, semantic communication, simulation software, successive-cancellation decoder.
\end{IEEEkeywords}
\section{Introduction}
\label{sec_intro}
Over the course of several decades, wireless communication has undergone a continuous evolution, driven by advancements in mathematical breakthroughs and novel innovations aimed at fulfilling the needs of human beings~\cite{BroadbandWhitePaper2020}. Consequently, the forthcoming wireless communication is expected to provide a highly sophisticated wireless experience, encompassing a vast range of implementations in various domains, such as extended reality (XR), self-sufficient robots, holoportation, and numerous other cutting-edge technologies~\cite{thushan2022terahertz}. Therefore, there is a necessity for a more sophisticated mode of communication.

The current communications paradigm has centered around transmitting bits while minimizing the occurrence of errors. This approach originated from Shannon's seminal 1948 paper~\cite{shannon_mathematical_1949}, which laid out the concept of \say{channel capacity}. In addition, it demonstrated that the rates below the channel capacity could be achieved without significantly increasing errors at the receiver's end. Researchers have pursued this topic for over five decades, eventually discovering capacity-achieving codes that work effectively over long block lengths.

The current receivers do not explicitly leverage the available source information at the transmitter side. Further, joint source-channel coding (JSCC)~\cite{504941} and unequal error protection (UEP)~\cite{6033770} have been extensively researched, but these were primarily focused on the transmitter side. However, the advent of modern image and video coding techniques has spurred a rising attraction to utilize artificial intelligence (AI) and machine learning (ML)~\cite{6gWhitepaper2020} for the efficient encoding of source information. This is further facilitated by the availability of image databases, which can be used to obtain style images for various scenarios. As a result, object classification is enhanced, leading to improved segmented images referred to as semantically coded images.

The idea behind semantic communication is to explore the knowledge base information at the receiver end to reduce communication overhead and enhance user experience. This concept is particularly significant in 6G and beyond~\cite{BroadbandWhitePaper2020}, given the substantial role played by Internet of Things (IoT) applications. In this context, the number of transmission bits and their abstract meaning holds paramount importance. However, achieving deep trustworthiness tailored to specific applications is more crucial than shallow precision at the bit level. Integrating semantic concepts with wireless communications presents several novel challenges~\cite{9398576}.

Polar codes \cite{Arikan, Rezaei2022, Rezaei20222,Rezaei20223,Combinational2023, 9621127, rezaei2023unrolled} are the first capacity-achieving error-correcting codes over binary-input discrete memoryless channels (B-DMC). They are constructed recursively using polarization phenomenon \cite{Arikan}, a feature that enables them to correct errors and optimize communication channels. As such, polar codes are a potential solution in the field of coding theory. The low-complexity encoding and decoding algorithms of polar codes have led to selection as the coding scheme for the control channel of enhanced mobile broadband (eMBB) in the fifth generation of new radio (5G-NR) wireless communication standards. 

In this paper, we propose a flexible software \cite{PythonCode} that automatically transmits the semantic segmentation map images over an additive white Gaussian noise (AWGN) channel using binary phase-shift keying (BPSK). This software is beneficial in investigating the effect of channel noise on end-to-end image communication systems utilizing semantic concepts. The well-known polar codes are chosen as the channel coding scheme, with the flexibility of selecting any code rate and code length. The dataset selected for the task is the popular common objects in context (COCO)-Stuff dataset \cite{cocostuff}, which is an augmented version of the COCO \cite{lin2014microsoft} dataset and contains $91$, different stuff classes. The output semantic map images corresponding to four different signal-to-noise ratios (SNRs) are generated to achieve a very small dataset, each containing a thousand images from the COCO-Stuff dataset \cite{PythonCode}. While this paper utilizes the COCO-Stuff dataset as an example, it should be noted that the software is not confined to this particular dataset and can be leveraged for transmitting images from any other dataset as well.

The remainder of the paper is organized as follows. In Section \ref{sec_back}, a background on end-to-end semantic communications, COCO-Stuff dataset, and polar codes will be provided. Section \ref{sec_post} presents the post-channel semantic map image generator software. The simulation results are summarized in Section \ref{sec_sim}, and finally, Section \ref{sec_conc} concludes this work.

\section{Background}
\label{sec_back}
\subsection{End-to-End Semantic Communications}
Currently, there is no comprehensive system model that integrates semantic and current communication systems for effective semantic communication. However, a preliminary system model for semantic communication has been proposed in~\cite{10100737}. According to this model, the source information is initially encoded with semantic coding schemes and then further encoded using current channel coding approaches before being transmitted via communication channels. The received bits are decoded using existing channel decoders at the receiver's end, and the semantic decoder produces the output. The success of this strategy relies heavily on the accuracy of the feature extraction process and its ability to meet the receiver's requirements. As such, there is a need for thorough research into feature extraction and optimization to ensure the effectiveness of semantic coding.

For semantic communication, the first essential step is semantic coding. This involves extracting and capturing the meaning or semantic features of the source and ensuring they align with the sufficient conditions of the receiver. For example, image/video applications segment images based on templates and use image databases such as the COCO dataset~\url{https://cocodataset.org}. However, defining a general framework for semantic coding is challenging since receiver requirements vary depending on the application. Task-oriented semantic extraction~\cite{10.1016/j.dsp.2021.103134} and coding can improve data rates significantly. Since there is not a single transmission system for semantic communication, designing a semantic communication system that aligns with the current communication framework is imperative.

The channel decoder initially decodes the semantic information from the received signal at the receiver end. The primary difficulty is guaranteeing that the transmitter's original semantic details are maintained during the communication. Subsequently, the disordered semantic information is fed as input to the semantic decoder, which then generates an output utilizing the existing knowledge base. For example, the authors in~\cite{10100737} propose a generative adversarial network (GAN) based semantic encoder, which produces the actual output image using the existing style image (knowledge base) and the received segmented map. 

\subsection{COCO-Stuff Dataset}
COCO~\cite{coco} is a popularly-used dataset in computer vision that serves as a benchmark for various image-based tasks. It is a comprehensive dataset that includes features such as object detection, segmentation, and captioning. This dataset has become a standard knowledge base for semantic communication-based image transmission systems due to its large-scale and diverse collection of images. COCO comprises over $118$K training images and $5$K in validation images containing various everyday objects captured in familiar settings. Additionally, the dataset features $1.5$ million object instances, $80$ object classes, and $91$ stuff classes, making it an extensive and varied dataset. Another unique feature of COCO is its inclusion of five captions per image and $ 250$K individuals with key points, making it a valuable resource for research in computer vision and artificial intelligence. Therefore, we investigate our analysis based on the COCO dataset. 

Numerous studies have utilized the COCO dataset to investigate various applications in image processing. One such study, named as CGBNet~\cite{8954873}, uses context encoding and multi-path decoding to create a semantic segmentation based on the COCO dataset. Another study employed a GAN-based image segmentation technique~\cite{8845685} which addresses the distribution similarity problem in image segmentation from natural language referring expressions. Additionally, researchers in~\cite{9281078} utilized the COCO stuff to investigate image coding strategies and develop a semantically structured bitstream to reduce complexity.

\subsection{Polar Codes}
Polar codes, invented by Arikan in $2009$ \cite{Arikan}, represent a distinctive class of Shannon's capacity-achieving error-correcting codes. Let us denote by $\mathcal{P}(N,K)$ a polar code of length $N \text{=} 2^n$, which contains $K$ information bits. The code rate then can simply be computed as $\mathcal{R} \text{=} {K/N}$. As the code length approaches infinity ($N \rightarrow \infty$), the polarization phenomenon allows for the physical channel to be divided into extremely reliable and unreliable virtual channels. The $K$ most reliable bit positions are included in the information set $\mathcal{I}$, while the remaining $N\text{-}K$ less reliable bit positions are included in the frozen set $\mathcal{F}$.

Mathematically speaking, binary polar codes also known as Arikan's codes, are a set of two bits to two bits transformation using a basic $2\times2$ polarization matrix known as binary kernel. The binary kernel is denoted by $G_2$ and defined as
\begin{equation}
    G_2 = 
      \left[\begin{array}{cc}
    1 & 1\\
    1 & 0
      \end{array}\right]. \qquad
      \label{eq_G2}
\end{equation}
By employing a linear transformation as $x = u\cdot G$, larger polar codes can be constructed in a recursive manner. Here $x$ denotes the encoded stream, $u$ represents an $N$-bit input vector and $G$ is the generator matrix created by the $n$-th Kronecker product matrix, i.e. 
\begin{equation}
G \triangleq T_{n_0}\otimes T_{n_1}\otimes...\otimes T_{n_s},
      \label{eq_G}
\end{equation}
where $T_{n_i}$s are squared kernel matrices. The input message is then integrated into the reliable bit positions of $u$ and the remaining bits of $u$ are set to zero. 

\section{Post-Channel Semantic Map Image Generator}
\label{sec_post}
\begin{figure*}
    \centering    \includegraphics[width=1.2\columnwidth]{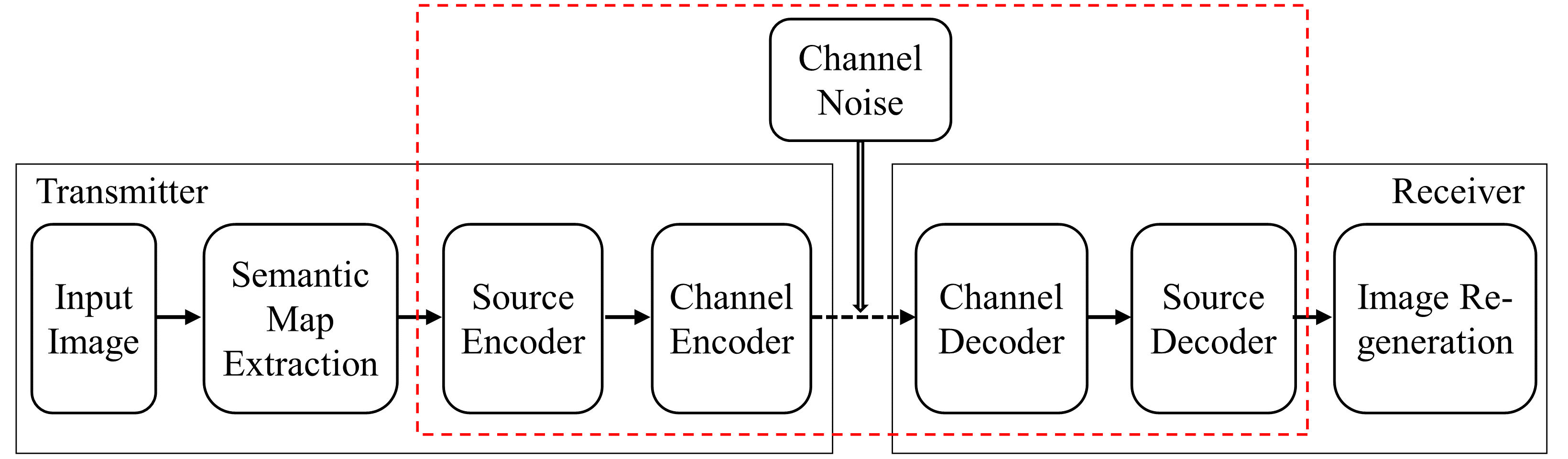}
    \caption{High-level architecture of an end-to-end image transmission system using semantic communications.}
    \label{fig:Gen_proc}
\end{figure*}

In this section, we will expound on the proposed post-channel semantic map image generator software. %Subsequently, we will delve into the impact of code length and code rate on the sample images across a range of various SNRs.
Fig. \ref{fig:Gen_proc} illustrates a high-level architecture of an end-to-end semantic-based image transmission system. A desired framework can be used to extract the semantic map images in the transmitter side. In this study, we utilize the COCO-Stuff dataset as it is readily available. The extracted semantic maps are subsequently subjected to encoding by polar codes, which have been employed as the channel (de)coder. The users have the flexibility to select the desired code length and code rate to achieve their objective. The encoded data is then transmitted through an AWGN channel using BPSK modulation, and the impact of channel noise on the image data is determined by the selected SNR. On the receiver side, a polar decoder is employed to decode the image data, and the resulting data is utilized to reconstruct the semantic map image. As expected, the quality of the regenerated image will be influenced by the channel noise. The proposed software is responsible for executing all the tasks delineated within the red dashed box.
Finally, log-likelihood ratios are used as the demaping method. The proposed software is scripted in Python and the specification of the channel is summarized in Table \ref{tab:chnlSpec}. 
\begin{table}
\centering
\caption{Channel (de)coding specification.}
\begin{tabular}{ll}
\hline
\textbf{Parameter} & \textbf{Value} \\\hline
Channel en/decoder& Polar Code\\
Length of information bits&Flexible  \\
Length of codeword& Flexible\\
Rate of Code& Flexible\\
Modulation scheme &BPSK \\
Number of bits per symbol & 2\\
Demaping method& Log-likelihood ratios\\
Channel Type& AWGN\\
\hline
\end{tabular}
\label{tab:chnlSpec}
\end{table}

%\hl{A comprehensive, step-by-step depiction of the software's execution sequence is outlined in Algorithm }\ref{Alg:SW}.
A comprehensive, step-by-step depiction of the software's execution sequence is outlined in Algorithm \ref{Alg:SW}, which provides a detailed flow to understand the concept of the execution. Moreover, Algorithm~\ref{Alg:decfunc} describes the channel decoder's function. Other than stated above, $D$ is the size of the dataset, $T_c$ is the CPU's elapsed time, \begin{math} {\mathbf{h_r}} \end{math} indicates reliable channels, and \begin{math} {\mathcal{CD}(.)} \end{math} denotes the channel decoder. Also, $\mathbf{I_{dd}}$ and $\mathbf{I_{ds}}$ represent the accumulated decoded data (all the decoded image data), and corresponding data of the decoded stream (only the latest decoded packet), respectively. 
\begin{algorithm}
\textbf{Inputs}:  $N$, $K$, $D$, $min\_SNR$, $max\_SNR$, $SNR\_step$\\
\textbf{Outputs}: Post-channel images, performance data, $T_c$\\

\textbf{Compute} $\mathbf{h_r}$\\
\For{$SNR \gets min\_SNR$ \textbf{to} $max\_SNR$ \textbf{by} $SNR\_step$}{   
    \For{$j \gets 1$ \textbf{to} $D$}{                   
            \textbf{read} $j$th image\\
            \textbf{reshape} image data and calculate $pixel\_count$\\
            \While{$pixel\_count > 0$}{ 
                 $\mathbf{msg} \gets$ next $K$ bits \\
                 $\mathbf{I_{ds}}$, $BER$, $FER \gets$ $\mathcal{CD}\left( N, K, \mathbf{msg}, \mathbf{h_r}, SNR \right)$\\
            $\mathbf{I_{dd}}$ $\gets$ $\mathbf{I_{dd}} + \mathbf{I_{ds}}$\\
            \textbf{write} image characteristics, $N$, $K$, $FER$, $BER$, $T_c$
            }
            \textbf{reshape} $\mathbf{I_{dd}}$ and construct an image \\
            \textbf{write} constructed image to the output folder          
        }}
\caption{Execution flow of the proposed software}
\label{Alg:SW}
\end{algorithm}
\begin{algorithm}
\textbf{Inputs}: $N$, $K$, $\mathbf{msg}$, $\mathbf{h_r}$, $SNR$\\
\textbf{Outputs}: $\mathbf{I_{ds}}$, $BER$, $FER$\\

\textbf{Insert} $\mathbf{msg}$ into reliable bit positions\\
\textbf{Encode} using polar codes\\
\textbf{Modulate} BPSK\\
\textbf{Pass} through an AWGN channel\\
\textbf{Decode} using polar decoder (estimate $\mathbf{I_{ds}}$)\\
\textbf{Compute} $BER$, $FER$\\
\textbf{Return} $\mathbf{I_{ds}}$, $BER$, $FER$
\caption{${\mathcal{CD}(.)}:$ Channel decoder's function }
\label{Alg:decfunc}
\end{algorithm}

\begin{figure*}
    \centering
    \includegraphics[width=1.2\columnwidth]{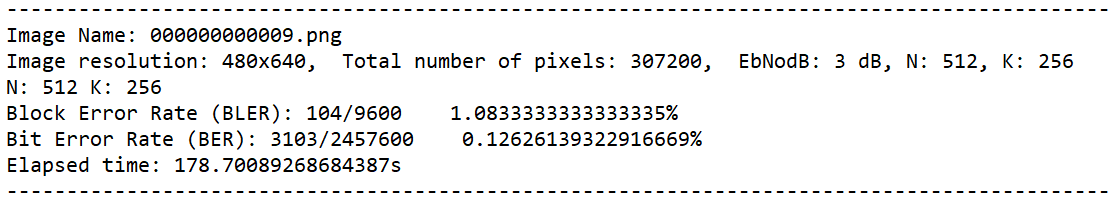}
    \caption{A text output generated by the software after transmitting a sample image.}
    \label{fig:sw_out}
\end{figure*}

The simulation software has the capability to transmit one or multiple images in one run. It generates the post-channel images along with a comprehensive text file that summarizes the image parameters, error rates, and channel specifications corresponding to all processed images. The proposed software's sample text output corresponding to a sample image is illustrated in Fig. \ref{fig:sw_out}. The report provides an overview of several key factors that impact image quality and processing efficiency. Specifically, the image resolution, number of pixels, SNR, polar code specifications, error rates (frame-error rate (FER) and bit-error rate (BER)), and CPU's elapsed time are all highlighted. Additionally, the study employs an AMD Ryzen $7$ PRO $5850$U x$64$ CPU operating at a frequency of $1.90$ GHz to execute the software. Notably, the CPU's elapsed time is significantly influenced by the resolution of the image. As a general rule, higher image resolutions result in longer elapsed times. The average time required to transfer an image from the COCO-Stuff dataset through the channel is roughly three minutes. This is substantiated by the fact that the sample image of Fig. \ref{fig:sw_out} necessitates the transmission of $300$K bits of data, which can be transferred through the use of $1200$ packets.
\section{Simulation Results}
\label{sec_sim}
\begin{figure*}[t]
    \centering
    \includegraphics[width=2\columnwidth]{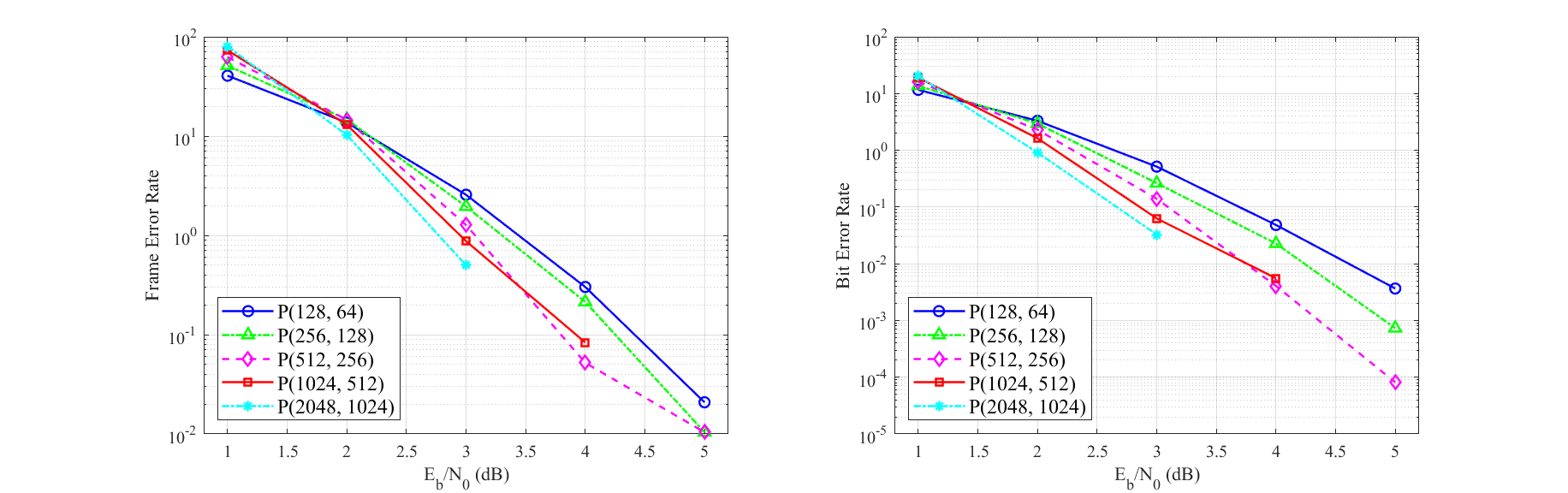}
    \caption{Effect of the block length of polar codes on the error-correction performance of a sample image over an AWGN channel.}
    \label{fig:ECC_variableN}
\end{figure*}

\begin{figure}
    \centering
    \includegraphics[width=1\columnwidth]{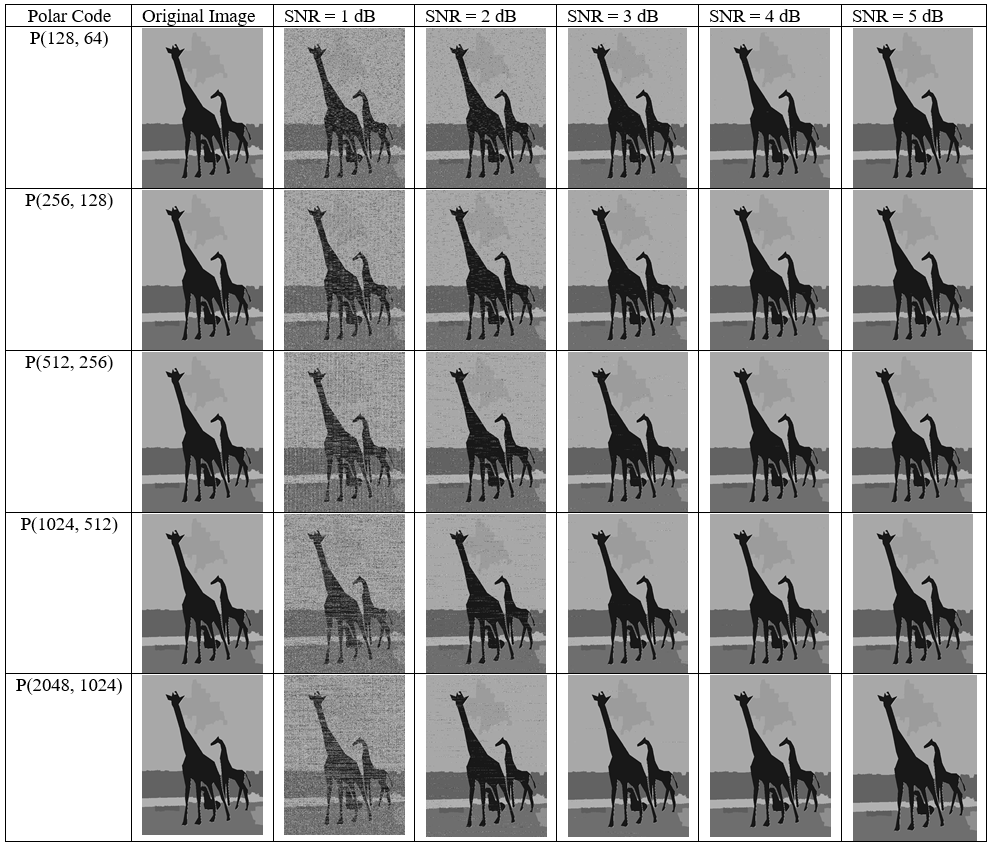}
    \caption{Effect of changing block length of polar codes with different SNRs on a sample image.}
    \label{fig:ImagesN}
\end{figure}

In this section, we will examine the effect of code length and code rate of polar codes on the error-correction performance. Fig.~\ref{fig:ECC_variableN} 
illustrates the impact of altering the code length on both FER and BER. All codes have a fixed rate of $\mathcal{R} = 1/2$. Fig. \ref{fig:ECC_variableN} demonstrates that increasing the code length results in superior FER and BER performances. This is due to higher polarization of larger polar codes meaning that some channels' reliability increases as the code length grows, while others decrease. As a result, we can choose channels with higher reliabilities while maintaining the same code rate. Fig. \ref{fig:ImagesN} depicts a sample image transmitted through a communication channel utilizing polar codes with varying lengths and a constant code rate of $\mathcal{R} = 1/2$. The figure illustrates that shorter codes perform better than longer codes when the SNR is $1$ dB. However, as the SNR increases, longer codes become more effective, which aligns with the findings presented in Fig. \ref{fig:ECC_variableN}. 

\begin{figure*}
    \centering
    \includegraphics[width=2\columnwidth]{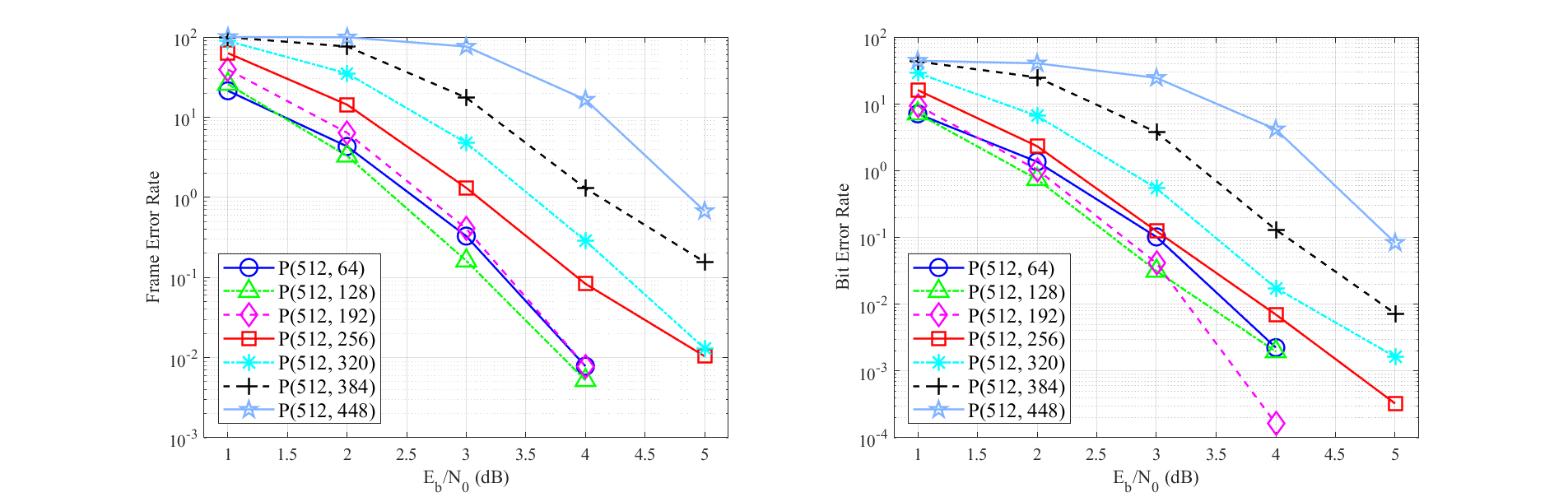}
    \caption{Effect of changing the rate of polar codes on the error-correction performance of a sample image over an AWGN channel.}
    \label{fig:ECC_variableK}
\end{figure*}

\begin{figure}
    \centering
    \includegraphics[width=1\columnwidth]{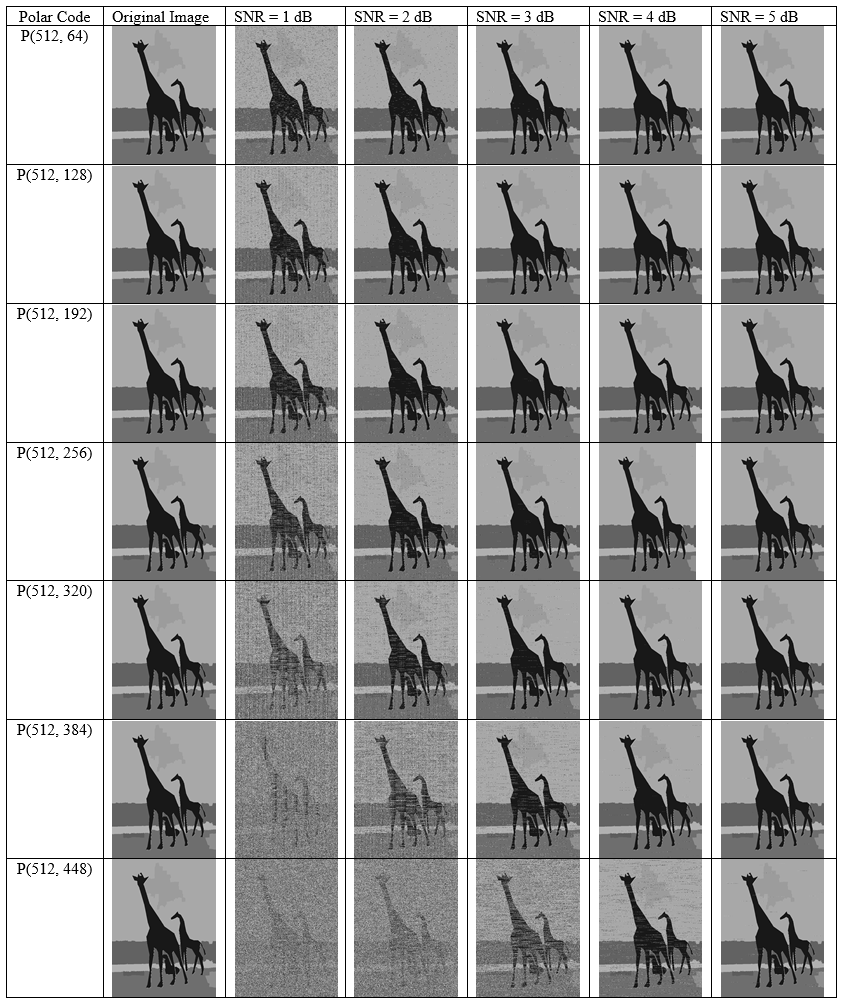}
    \caption{Effect of changing the rate of polar codes with different SNRs on a sample image.}
    \label{fig:ImagesK}
\end{figure}

The impact of altering the code rate is depicted in Fig. \ref{fig:ECC_variableK}. It is evident that as the code rate increases, the FER and BER tend to increase.
 This is because less reliable virtual channels are used to transmit data as the code rate increases. Fig. \ref{fig:ImagesK} displays a sample image transmitted through the channel utilizing a polar code of size $N=512$ and various code rates. It is apparent that images transmitted with a high code rate and low SNR exhibit the highest amount of noise.
 \section{Conclusion}
\label{sec_conc}
Flexible simulation software that automatically transmits semantic segmentation map images using polar codes is presented in this paper. The proposed software allows for a comprehensive analysis of the impact of channel noise on semantic map images within end-to-end image transmission systems. While the COCO-Stuff dataset is selected in this paper, it is essential to note that the software can transmit images from any preferred dataset. Moreover, the user is also empowered to choose the desired coding parameters and signal-to-noise ratio, enhancing the software's flexibility and usability. With its advanced features and adaptability, this simulation software represents a significant step forward in the field of semantic image transmission with wireless communication. 

\section*{Acknowledgment}
This research has been supported by the Academy of Finland, 6G Flagship program under Grant 346208.


\end{document}